\documentclass[12pt]{article}

\usepackage{graphicx}
\usepackage{scicite}

\usepackage{times}

\def\bea{\begin{eqnarray}}      \def\eea{\end{eqnarray}}

\newenvironment{sciabstract}{%
\begin{quote} \bf}
{\end{quote}}

\newcounter{lastnote}
\newenvironment{scilastnote}{%
\setcounter{lastnote}{\value{enumiv}}%
\addtocounter{lastnote}{+1}%
\begin{list}%
{\arabic{lastnote}.} {\setlength{\leftmargin}{.22in}}
{\setlength{\labelsep}{.5em}}} {\end{list}}

\title{Quantum Spin Hall Insulator State in HgTe Quantum Wells}

\author
{Markus K\"{o}nig$^{1}$, Steffen Wiedmann$^{1}$, Christoph Br\"{une}$^{1}$, Andreas Roth$^{1}$,\\ Hartmut Buhmann$^{1}$, Laurens W. Molenkamp$^{1,\ast}$, \\Xiao-Liang Qi$^{2}$ and Shou-Cheng Zhang$^{2}$\\
\\
\normalsize{$^{1}$Physikalisches Institut (EP III), Universit$\ddot{\rm a}$t W$\ddot{\rm u}$rzburg}\\
\normalsize{D-97074 W$\ddot{\rm u}$rzburg, Germany}
\\
\normalsize{$^{2}$Department of Physics, McCullough Building,
Stanford
University}\\
\normalsize{Stanford, CA 94305-4045}\\
\normalsize{$^\ast$To whom correspondence should be addressed:}\\
\normalsize{molenkmp@physik.uni-wuerzburg.de}
}

\date{}

\begin{document}


\baselineskip24pt


\maketitle


\begin{sciabstract}
Recent theory predicted that the Quantum Spin Hall Effect, a fundamentally
novel quantum state of matter that exists at zero external
magnetic field, may be realized in HgTe/(Hg,Cd)Te quantum wells.
We have fabricated such
sample structures with low density and high mobility in which we can tune,
through an external gate voltage, the carrier conduction from
$n$-type to the $p$-type, passing through an insulating regime. For
thin quantum wells with well width $d< 6.3$ nm, the
insulating regime shows the conventional behavior of vanishingly
small conductance at low temperature. However, for thicker quantum
wells ($d> 6.3$ nm), the nominally insulating
regime shows a plateau of residual conductance close to $2e^2/h$.
The residual conductance is independent of the sample width,
indicating that it is caused by edge states. Furthermore, the
residual conductance is destroyed by a small external magnetic
field. The quantum phase transition at the critical thickness, $d= 6.3$ nm, is also
independently determined from the magnetic field induced insulator
to metal transition. These observations provide
experimental evidence of the quantum spin Hall effect.
\end{sciabstract}


\newpage

The theoretical prediction of the intrinsic spin Hall effect in
metals and insulators\cite{murakami2003,sinova2004,murakami2004a}
has generated great interests in the field of spintronics, since
this effect allows for direct electric manipulation of the spin
degrees of freedom without a magnetic field, and the resulting
spin current can flow without dissipation. These properties could
lead to promising spintronic devices with low power dissipation.

However, beyond the potential technological applications, the
intrinsic spin Hall effect has guided us in the search for new and
topologically non-trivial states of matter. The quantum Hall state
gives the first, and so far the only example of a topologically
non-trivial state of matter, where the quantization of the Hall
conductance is protected by a topological invariant. The quantum
spin Hall (QSH)
insulators\cite{kane2005A,bernevig2006A,bernevig2006d} have a
similar, but distinct non-trivial topological property. The QSH
insulators are invariant under time reversal, have a charge
excitation gap in the bulk, but have topologically protected
gapless edge states that lie inside the bulk insulating gap. This
type of insulator is typically realized in spin-orbit coupled
systems; the corresponding edge states have a distinct helical
property: two states with opposite spin-polarization
counter-propagate at a given edge\cite{kane2005A,wu2006,xu2006}.
The edge states come in Kramers' doublets, and time reversal
symmetry ensures the crossing of their energy levels at special
points in the Brillouin zone. Because of this energy level
crossing, the spectrum of a QSH insulator cannot be adiabatically
deformed into that of a topologically trivial insulator without
helical edge states; therefore, in this precise sense, the QSH
insulators represent a topologically distinct new state of matter.

It has been proposed theoretically that HgTe/(Hg,Cd)Te quantum
wells provide a natural realization of the quantum spin Hall
effect\cite{bernevig2006d}. In zincblende-type semiconductor
quantum wells, there are four relevant bands close to the Fermi
level. The $E1$ band basically consists of the two spin states of
the $s$ orbital, while the $HH1$ band basically consists of the
$|p_x+ip_y, \uparrow\rangle$ and $|-(p_x-ip_y), \downarrow\rangle$
orbitals. The effective Hamiltonian near the $\Gamma$ point is
given by
\begin{equation}
H_{eff}(k_x,k_y)=\left(\matrix{ H(k)&0 \cr 0&H^*(-k)} \right), \ \
H=\epsilon(k) + d_i(k) \sigma_i
 \label{dirac}
\end{equation}\noindent
where $\sigma_i$ are the Pauli matrices, and
\begin{eqnarray}
& d_1+id_2=A(k_x+i k_y)\equiv Ak_+ \nonumber \\
& d_3 = M - B (k_{x}^2+k_{y}^2) \ , \ \epsilon_k = C -
D(k_{x}^2+k_{y}^2). \label{d-expansion}
\end{eqnarray}\noindent
Here, $k_x$ and $k_y$ are momenta in the plane of the 2DEG, while
$A,B,C,D$ are material specific constants. Spin-orbit coupling is
naturally built-in in this Hamiltonian through the spin-orbit
coupled $p$ orbitals $|p_x+ip_y, \uparrow\rangle$ and
$|-(p_x-ip_y), \downarrow\rangle$. Two dimensional materials can
be grouped into three types according to the sign of the Dirac
mass parameter $M$. In conventional semiconductors such as GaAs
and CdTe, the $s$-like $E1$ band lies above the $p$-like $HH1$
band, and the mass parameter $M$ is positive. Semi-metals such as
graphene\cite{Novoselov2005,Zhang2005} are described by a massless 
Dirac model with $M=0$, although the bands have a different physical 
interpretation. In so-called ``inverted" semiconductors such as HgTe, the $s$-like
orbital lies below the $p$-like orbitals; therefore, the Dirac
mass parameter $M$ in the HgTe/(Hg,Cd)Te quantum wells can be
continuously tuned from a positive value $M>0$ for thin quantum
wells with thickness $d<d_c$ to a negative value $M<0$ for thick
quantum wells with $d>d_c$. A topological quantum phase transition
occurs at $d=d_c$, where the system is effectively described by
a massless Dirac theory just like for graphene. The nature of
this quantum phase transition has also been investigated in more
realistic models beyond the simple four band model presented here,
reaching the same conclusion\cite{dai2007}. 

The QSH phase occurs in the inverted regime where $M<0$, i.e.,
when $d>d_c$. The sample edge can be viewed as a
domain wall of the mass parameter $M$, separating the
topologically non-trivial phase with $M<0$ from the topologically
trivial phase with $M>0$, which is adiabatically connected to the
vacuum \cite{PbTe}. Massless helical states are confined on the sample edge.
The sample has a finite conductance even when the
Fermi level lies inside the bulk insulating gap. Therefore, as
suggested in Ref. \citen{bernevig2006d}, the QSH state can be
experimentally detected by measuring a residual conductance
plateau as one varies the gate voltage in the nominally insulating
regime. Furthermore, because the current is carried by the edge
states, the conductance should be independent of sample width.
Protected by the time reversal symmetry, non-magnetic impurities
or any other time-reversal invariant local perturbations cannot
cause elastic backscattering of the helical edge states, which
warrants the topological robustness of the edge state conductance.
However, the presence of magnetic field breaks time reversal symmetry,
therefore can open up a gap in the energy spectrum of the edge states,
and remove the residual conductance due to the edge states.

We set out to test these theoretical predictions by
measuring the transport properties of HgTe/(Hg,Cd)Te quantum
wells as a function of the sample thickness, the gate voltage and
the external magnetic field. We use
modulation-doped type-III\cite{novik2005}
HgTe/Hg$_{0.3}$Cd$_{0.7}$Te quantum well (QW) structures
fabricated by molecular beam epitaxy\cite{somegrowthpaper}, with
widths\cite{remarkonKumpf} varying from 5.5 nm ($d<d_c$) to 12 nm
($d>d_c$). Dedicated low-thermal budget optical and e-beam
lithography is used to structure devices in Hall bar geometry with
dimensions, $(L\times W)$, of  $(600 \times 200)$~$\mu$m$^2$,
$(20.0 \times 13.3)$~$\mu$m$^2$, $(1.0 \times 1.0)$~$\mu$m$^2$,
and $(1.0 \times 0.5)$ $\mu$m$^2$ (see inset in Fig.\ 1A). All
devices have a 110 nm thick Si$_3$N$_4$/SiO$_2$
multilayer gate insulator\cite{hinze2006} and a 5/30 nm Ti/Au gate
electrode stack. Transport measurements are done in a
$^3$He/$^4$He-dilution refrigerator (base temperature $T < 30$ mK,
uniaxial fields up to 18 T) and in a $^4$He cryostat fitted with a
vector magnet ($T= 1.4$ K, and fields up to 300 mT in variable
direction), using lock-in techniques. At zero gate voltage, the
samples are $n$-type, exhibiting carrier densities between $1.3 \times 10^{11}$ cm$^{-2}$
and $3.5 \times 10^{11}$ cm$^{-2}$ and mobilities up to $1.5
\times 10^5$ cm$^{2}$/(Vs). The carrier density can be reduced
continuously by applying a negative gate voltage to the Au
electrode with respect to carriers in the QW. The Si-O-N gate
insulator stack allows for quite large gate voltages, enabling us
to gate the samples through the gap from $n$-type to
$p$-type conductance.

The change in carrier type can be monitored from Hall
experiments, as shown in Fig.\ 1A for a large $\left[(L\times W) =
(600 \times 200)\: \mu{\rm m}^2 \right]$ Hall bar with a well width of 6.5
nm, at 30 mK. The change in carrier type is directly reflected in a
sign change of the slope of the Hall resistance $R_{xy}$, and we
can directly infer that the carrier density varies from $n= 1.2
\times 10^{11}$~cm$^{-2}$ at gate voltage $V_g =$ -1.0 V to $p =
1.0 \times 10^{10}$~cm$^{-2}$ at $V_g =$ -2.0~V. At modest
magnetic fields, for both $n$-type and $p$-type channels, $R_{xy}$
exhibits quantum Hall plateaus, indicative of the good quality of
the material, until at fields above ca. 3 T the last Landau level is
depleted.

Remarkable transport behavior is observed for -1.9 V $ \leq V_g \leq $
-1.4 V (see in particular the green and the red traces in Fig.\ 1),
where the sample is insulating at zero magnetic field
(i.e., the Fermi level is in the gap). For these gate voltages, we
observe that the sample undergoes a phase transition from an
insulating state to a QH state with a quantized Hall conductance
of $G_{xy}=\pm e^2/h$, either $n$- or $p$-type depending on $V_g$, at a
small (order 1-2 T) applied magnetic field. The sample remains
in the QH state for a few more T, and then becomes once again insulating.
We have observed this phenomenon in a number of samples in the
inverted regime, with $6.5$ nm $< d < 12$ nm.

The phase transition from an insulating state to a QH state is a
non-trivial consequence of the inverted band structure of the QSH
insulator, and can be explained by the level crossing of the $E1$
and $HH1$ Landau levels, which can be directly obtained from the
minimal coupling of the simple Dirac model Eq.\ \ref{dirac} to a
perpendicular magnetic field $B_\perp$. If we only consider the
orbital effects of the magnetic field, two series of Landau levels
are obtained from the upper and lower $2\times 2$ blocks of
Hamiltonian, Eq.\ \ref{dirac}, in which the two levels closest to the
Fermi energy are given by $E_+=C+M-(D+B)l_c^{-2}$ and
$E_-=C-M-(D-B)l_c^{-2}$, with $l_c\equiv \sqrt{  \hbar
 / ({eB_\perp})}$. Thus the condition for level crossing is given by
$E_+-E_-=2M-2Bl_c^{-2}=0$ or $B^c_\perp=(\hbar M) /
({eB})$. Generally, the $B$ parameter is always
negative, therefore, we can see that the level crossing occurs
only in the inverted region with $M<0$.

The Landau levels for the normal ($d<d_c$) and the inverted ($d>d_c$) regime
are shown in Fig.\ 2A and B, respectively. Edge states in the presence of
an external magnetic field can be easily obtained by solving our
simple Dirac model in Eq.\ \ref{dirac} with open boundary condition
along one direction and periodic boundary condition along the
other direction. Fig.\ 2C shows the bulk and edge states for a
conventional insulator. With increasing thickness $d$, the two
states closest to the Fermi energy approach and then cross each
other. This ``band inversion"  leads to the bulk and edge states
[Fig.\ 2D]. The Fermi energy crosses a pair
of counter-propagating edge states on each edge, resulting in no
net Hall effect. These counter-propagating edge states are similar
to the helical edge states of the QSH insulator. However, due to
the presence of the magnetic field and the breaking of the time
reversal symmetry, they are not robustly protected. Starting from
this case, increasing magnetic field shifts the red Landau level towards
higher and the blue one towards lower energies.
When one of the bulk Landau level crosses the Fermi level, there is
a single edge state on each edge left, and one obtains a net QH
effect with either $G_{xy}=e^2/h$ [Fig.\ 2E], or
with $G_{xy}=-e^2/h$ [Fig.\ 2F]. When the magnetic
field is increased further, the second bulk Landau level crosses
the Fermi level, and one reaches the conventional
insulator case [Fig.\ 2C] but with the colors of the Landau level interchanged.
In models with bulk inversion asymmetry (BIA)\cite{dai2007}, the level crossing between $E1$
and $HH1$ Landau levels at $B^c_\perp$ can be avoided, and the phase
regions (i) and (ii) in Fig.\ 2B become connected. Generally, the
non-monotonic dependence of the Landau level energies leads to the
transition from the insulating state to the QH state at a
constant Fermi energy, when the magnetic field is varied.

While the four band Dirac model (Eq.\ \ref{dirac}) gives a simple
qualitative understanding of this novel phase transition, we have
also performed more realistic and self-consistent eight band
$\textbf{k} \cdot \textbf{p}$ model calculations\cite{novik2005}
for a $6.5$ nm quantum well, with the fan chart of the Landau levels
displayed in Fig.\ 1B. The two anomalous Landau levels cross at a
critical magnetic field $B^c_\perp$, which evidently depends on well
width. This implies that when a sample has its Fermi energy in the
gap at zero magnetic field, this energy will always be crossed by
the two anomalous Landau levels, resulting in a QH
plateau in between the two crossing fields. Fig.\ 3 summarizes the
dependence of $B^c_\perp$ on well width $d$. The open red squares are
experimental data points that result from fitting the 8-band
$\textbf{k} \cdot \textbf{p}$ model to experimental data as in
Fig.\ 1, while the filled red triangles solely result from the
$\textbf{k} \cdot \textbf{p}$ calculation. For reference, the
calculated gap energies are also plotted in this graph
as open blue circles. The band inversion is reflected in the sign
change of the gap. Note that for relatively wide wells ($d > 8.5$
nm) the (inverted) gap starts to decrease in magnitude. This is
because for these well widths, the band gap no longer occurs
between the $E1$ and $HH1$ levels, but rather between $HH1$ and
$HH2$ - the second confined hole-like level, as schematically
shown in the inset of Fig.\ 3 [see also \cite{pfeuffer-jeschke2000}].
Also in this regime, a band
crossing of conductance- ($HH1$) and valence- ($HH2$) band derived
Landau levels occurs with increasing magnetic
field\cite{novik2005, pfeuffer-jeschke2000, schultz1998}. Fig.\ 3
clearly illustrates the quantum phase transition that
occurs as a function of $d$ in the HgTe QWs: only for $d>d_c$,
$B^c_\perp$ exists, and at the same time the energy gap is negative
(i.e., the band structure is inverted). The experimental data allow
for a quite accurate determination of the critical thickness,
yielding $ d_c = $ 6.3 $\pm$ 0.1 nm.

The actual existence of edge channels in insulating inverted QWs
is only revealed when studying smaller Hall bars (note that the
typical mobility of $10^5$ cm$^{2}$/(Vs) in $n$-type material implies
an elastic mean free path of $l_{\rm mfp} \approx 1$ $\mu$m\cite{daumer, koenig06} - and one may
anticipate lower mobilities in the nominally insulating regime).
The pertinent data is shown in Fig.\ 4, which plots the zero
$B$-field four terminal resistance $R_{14,23}\equiv V_{23}/I_{14}$
as a function of normalized gate voltage ($V_{thr}$ is defined as
the voltage for which the resistance is largest) for several
devices, that are representative for the large number of
structures we have investigated. $R_{14,23}$ is measured while
the Fermi level in the device is scanned through the gap. In the
low resistance regions at positive $V_g-V_{thr}$ the sample is
$n$-type, at negative $V_g-V_{thr}$ the sample is $p$-type.

The black curve labeled I in Fig.\ 4 is obtained from a medium
sized [$(20.0 \times 13.3)$ $\mu$m$^2$] device with 5.5 nm QW, and
shows the behavior we observe for all devices with a normal band
structure: when the Fermi level is in the gap, $R_{14,23}$
increases strongly and is at least several tens of M$\Omega$ (this
actually is the noise floor of the lock-in equipment used in the
experiment). This clearly is the expected behavior for a
conventional insulator.  However, for all devices containing an
inverted QW, the resistance in the insulating regime remains
finite. $R_{14,23}$  plateaus out at well below 100 k$\Omega$
(i.e., $G_{14,23}=0.3 e^2/h$) for the blue curve labeled II,
which is again for a $(20.0 \times 13.3)$ $\mu$m$^2$ device
fabricated by optical lithography, but now contains a 7.3 nm wide
QW. For much shorter samples ($L=1.0$ $\mu$m, green and red
curves III and IV) fabricated from the same wafer, $G_{14,23}$
actually reaches the predicted value close to $2e^2/h$. This observation
provides firm evidence for the existence of the quantum spin Hall
insulator state for inverted HgTe QW structures.

In Fig.\ 4, we have included data on two $d=7.3$ nm, $L=1.0$ $\mu$m
devices. The green trace (III) is from a device with $W=1.0$ $\mu$m,
the red trace (IV) corresponds to $W=0.5$ $\mu$m. Clearly, the
residual resistance of the devices does not depend on the width of
the structure, which is evidence that the transport occurs through
edge channels\cite{longer}. One notices that the traces for the
$d=7.3$ nm, $L=1.0$ $\mu$m devices do not reach all the way into
the $p$-region. This is because the electron-beam lithography needed
to fabricate the devices increases the intrinsic ($V_g=$ 0 V)
carrier concentration. In addition, one notices fluctuations on
the conductance plateaus in traces II, III, and IV. These
fluctuations are actually reproducible and do not stem from, e.g.,
electrical noise.
While all $R_{14,23}$ traces discussed so far were taken at base temperature
(30 mK) of our dilution refrigerator, the conductance plateaus are
not at all limited to this very low temperature regime. In the
inset of Fig.\ 4 we reproduce the green 30 mK trace III on a linear
scale, and compare it with a trace (in black) taken at 1.8 K from
another $(L \times W)=(1.0 \times1.0)$~$\mu$m$^2$ sample, that was fabricated
from the same wafer. In the fabrication of this sample, we used a
lower illumination dose in the e-beam lithography, resulting in a
better (but still not quite complete) coverage of the $n$-$i$-$p$
transition. Clearly, in this further sample, and at 1.8 K, the $2
e^2/h$ conductance plateau is again present, including (thermally
smeared) conductance fluctuations.

In the pure two terminal geometry, with only source and drain contacts
(contacts 1 and 4, inset Fig 1A), the
two counter propagating helical edge states at one given edge
connect the chemical potential from the source and drain,
respectively, and they are not in equilibrium with each other
because the elastic backscattering vanishes between these two
channels. In the absence of voltage probes 2,3,5, and 6, as indicated in
the inset of Fig.\ 1, the two terminal conductance should give
$2e^2/h$. Now we consider the presence of the voltage probes.
Since our voltage probes are not spin sensitive, the voltage
measurement necessarily leads to the equilibration of the two
helical channels with the opposite spin orientation. A simple
Landauer-B\"uttiker type of calculation shows that the four terminal
resistance should in fact be given by $R_{14,23}=h/2e^2$.
In the presence of the voltage probes, the
voltage drops $V_{12}$, $V_{23}$ and $V_{34}$ add in series to
give a higher resistance of $R_{14}\equiv V_{14}/I_{14}=3h/2e^2$.
These results are valid as long as the distance between the
voltage probes is less than the inelastic mean free path $l_{in}$.
While elastic scatterers can not cause backscattering of the helical
edge states, inelastic scatterers can. We estimate the inelastic
mean free path to be $l_{in} > 1 \mu m$ at our measurement
temperature. Therefore, for the large sample (II), where the
distance between the voltage probes exceeds the inelastic mean
free path $l_{in}$, we expect the residual resistance to be
higher, consistent with the experimental measurement shown in the
trace (II) in Fig.\ 4.

Another intriguing observation is that the QSH effect is destroyed
by applying only a small magnetic field
perpendicular to the 2DEG plane. Fig.\ 5 shows that the
magnetoresistance is actually strongly anisotropic. (This data has
been obtained in the vector magnet system at 1.4 K.) A very sharp, cusp-like
conductance peak is observed for perpendicular
field, with the full width half-maximum (FWHM) $B_\perp^{FWHM}\simeq 28 $ mT\cite{noteaboutTdependence}.
The peak broadens strongly when the
magnetic field is tilted into the QW plane. For fully in-plane
fields, the QSH conductance can be observed over a much wider
magnetic field range ($B_{\parallel}^{\rm FWHM} \approx 0.7$ T).

The robustness of the helical edge states is ensured by the
time-reversal symmetry. A magnetic field breaks time reversal
symmetry, and thus turns on a gap between the two otherwise
degenerate helical edge states. The perpendicular and in-plane magnetic
field lead to different gaps, depending on the matrix elements of
the magnetization operator: \\
$E_{\rm gap\perp}=\left\langle
\uparrow\right|\left({\bf
\hat{z}\cdot\hat{r}\times\hat{j}}+g_\perp\mu_B{
S}_{\perp}\right)\left|\downarrow\right\rangle\left|B\right|$ and
$E_{\rm gap\parallel}=\left\langle \uparrow\right|g_\parallel
\mu_BS_\parallel\left|\downarrow\right\rangle\left|B\right|$, in
which $\hat{\bf r},\hat{\bf j}$ are the position and electric
current operator, respectively, and $\hat{\bf z}$ is the unit-vector
perpendicular to the quantum well plane. $S_{\perp(\parallel)}$
stands for the dimensionless part of the Zeeman-coupling matrix
element in perpendicular (parallel) magnetic field. We can estimate
the magnitude of these two gaps by noting that $\left\langle
\uparrow\right|{\bf
\hat{z}\cdot\hat{r}\times\hat{j}}\left|\downarrow\right\rangle\sim
ev\xi$ and
$\left\langle\uparrow\right|S_{\parallel,\perp}\left|\downarrow\right\rangle\sim
1$, in which $v$ and $\xi$ are the Fermi velocity and width of the
edge channels, respectively. $v$ and $\xi$ can be obtained from the
Dirac parameters as $v=A/\hbar$ and $\xi\simeq\hbar
v/\left|M\right|$. The parameters for the $d=7.3$ nm quantum well give
the dimensionless ratio $ev\xi/\mu_B\sim 280$, which thus leads to
$E_{\rm gap\perp}/E_{\rm gap\parallel}\sim 10^2$. From this
estimate, we can see that the strong anisotropy observed in the
experiments originates from the high Fermi velocity of the edge
states and the small bulk gap $M$, which together make the orbital
magnetization dominant.

So far our experiments have only measured the charge transport
properties. While the QSH effect manifests itself already in the
change in transport properties, we would still like to
experimentally confirm the spin accumulation resulting from the spin
Hall effect\cite{kato2004,wunderlich2005} in the topologically
non-trivial insulating regime, and compare both electric and
magnetic results with the experiments of the spin Hall effect in
the metallic regime. It would also be interesting to explore the
regime close to the quantum phase transition point of $d=d_c$, and
compare the transport properties with that of the recently
discovered graphene. In many ways, the HgTe quantum well system can be
viewed as a tunable graphene system, where the Dirac mass term can
be tuned continuously to zero from either the positive
(topologically trivial) or the negative (topologically
non-trivial) side.

\newpage

\bibliography{QSHI_cm}

\begin{thebibliography}{10}

\bibitem{murakami2003}
S.~Murakami, N.~Nagaosa, \textrm{S.C. Zhang}, {\it Science\/} {\bf 301}, 1348
  (2003).

\bibitem{sinova2004}
\textrm{J. Sinova}, {\it et~al.\/}, {\it Phys. Rev. Lett.\/} {\bf 92}, 126603
  (2004).

\bibitem{murakami2004a}
S.~Murakami, N.~Nagaosa, \textrm{S.C. Zhang}, {\it Phys. Rev. Lett.\/} {\bf
  93}, 156804 (2004).

\bibitem{kane2005A}
\textrm{C. L. Kane}, \textrm{E. J. Mele}, {\it Phys. Rev. Lett.\/} {\bf 95},
  146802 (2005).

\bibitem{bernevig2006A}
\textrm{B.A. Bernevig}, \textrm{S.C. Zhang}, {\it Phys. Rev. Lett.\/} {\bf 96},
  106802 (2006).

\bibitem{bernevig2006d}
\textrm{B. A. Bernevig}, \textrm{T. L. Hughes}, \textrm{S.C. Zhang}, {\it
  Science\/} {\bf 314}, 1757 (2006).

\bibitem{wu2006}
\textrm{C. Wu}, \textrm{B.A. Bernevig}, \textrm{S.C. Zhang}, {\it Phys. Rev.
  Lett.\/} {\bf 96}, 106401 (2006).

\bibitem{xu2006}
\textrm{C. Xu}, \textrm{J. Moore}, {\it Phys. Rev. B\/} {\bf 73}, 045322
  (2006).

\bibitem{Novoselov2005}
K.~S. Novoselov, {\it et~al.\/}, {\it Nature\/} {\bf 438}, 197 (2005).

\bibitem{Zhang2005}
Y.~Zhang, Y.~Tan, H.~L. Stormer, P.~Kim, {\it Nature\/} {\bf 438}, 201 (2005).

\bibitem{dai2007}
X.~Dai, T.~L. Hughes, X.~Qi, Z.~Fang, S.~Zhang, arxiv: cond-mat/0705.1516.

\bibitem{PbTe}
Similar mass domain walls have been proposed to occur in three-dimensional in
  PbTe/(Pb,Sn)Te heterostructures, where the $L^\pm_6$ bands change position as
  a function of Sn concentration \cite{Volkov1985,Fradkin1986}.

\bibitem{novik2005}
E.~G. Novik, {\it et~al.\/}, {\it Phys. Rev. B\/} {\bf 72}, 035321 (2005).

\bibitem{somegrowthpaper}
C.~Becker, {\it et~al.\/}, {\it Phys. Stat. Sol. (c)\/} {\bf 4}, 3382 (2007).

\bibitem{remarkonKumpf}
Well thicknesses have been calibrated using X-Ray Reflectivity measurements at
  the DESY synchrotron in Hamburg, Germany.

\bibitem{hinze2006}
J.~Hinz, {\it et~al.\/}, {\it Semicond. Sci. Technol.\/} {\bf 21}, 501 (2006).

\bibitem{pfeuffer-jeschke2000}
A.~Pfeuffer-Jeschke, Ph.D. Thesis, University of W{\"u}rzburg, Germany, 2000.

\bibitem{schultz1998}
M.~Schultz, {\it et~al.\/}, {\it Phys. Rev. B\/} {\bf 57}, 14772 (1998).

\bibitem{daumer}
V.~Daumer, {\it et~al.\/}, {\it Appl. Phys. Lett.\/} {\bf 83}, 1376 (2003).

\bibitem{koenig06}
M.~K\"onig, {\it et~al.\/}, {\it Phys. Rev. Lett.\/} {\bf 96}, 76804 (2006).

\bibitem{longer}
We have actually observed a similar independence of resistance on sample width
  in the $d=$7 nm, $L=20.0$ $\mu$m devices, showing that also in these larger
  structures the conductance is totally dominated by edge channels.

\bibitem{noteaboutTdependence}
FWHM of the magnetoresistance peak is about 10 mT at 30 mK, increasing to 28 mT
  at 1.4 K.

\bibitem{kato2004}
Y.~K. Kato, R.~C. Myers, A.~C. Gossard, D.~D. Awschalom, {\it Science\/} {\bf
  306}, 1910 (2004).

\bibitem{wunderlich2005}
J.~Wunderlich, B.~Kaestner, J.~Sinova, T.~Jungwirth, {\it Phys. Rev. Lett.\/}
  {\bf 94}, 047204 (2005).

\bibitem{Volkov1985}
B.~A. Volkov, O.~A. Pankratov, {\it JETP Lett.\/} {\bf 42}, 178 (1985).

\bibitem{Fradkin1986}
E.~Fradkin, E.~Dagotto, D.~Boyanovsky, {\it Phys. Rev. Lett.\/} {\bf 57}, 2967
  (1986).

\end{thebibliography}

\bibliographystyle{Science}

\begin{scilastnote}
\item We wish to thank A. Bernevig, X. Dai, Z. Fang, T. Hughes, C.
X. Liu and C.J. Wu for insightful discussions, C.R. Becker and V.
Hock for sample preparation, and C. Kumpf for calibrating the well
widths of the HgTe samples. This work is supported by the DFG (SFB
410), by the German-Israeli Foundation for Scientific Research and
Development (Grant No.881/05), by NSF through the grants
DMR-0342832, and by the US Department of Energy, Office of Basic
Energy Sciences under contract DE-AC03-76SF00515, and Focus Center
Research Program (FCRP) Center on Functional Engineered
Nanoarchitectonics (FENA).

\end{scilastnote}

\newpage

\section{Figure Captions}

\textbf{Fig. 1} (A) Hall resistance, $R_{xy}$, of a $(L\times W) = (600\times 200)$ $\mu$m$^2$ QW structure with 6.5 nm well width for different carrier concentrations obtained for gate voltages $V_g$ of -1.0 V (black), -1.1 V (purple), -1.2 V (navy), -1.3 V (blue), -1.35 V (cyan), -1.4 V (green), -1.7 V (red), -1.8 V (orange), -1.9 V (brown), and -2.0 V (black, lower curve). For decreasing $V_g$, the $n$-type carrier concentration decreases and a transition to a $p$-type conductor is observed, passing through an insulating regime  between -1.4 and -1.9~V at $B=0$~T.
The inset shows a schematic sample layout with ohmic contacts labeled 1 to 6. The grey shaded region indicates the top gate electrode and defines the part of the sample where the carrier concentration and type can be changed. Red and blue arrows indicate the counter-propagating spin-polarized edge channels of the QSH effect.
(B) The Landau level fan chart of a 6.5~nm quantum well obtained from an eight band $\textbf{k}\cdot \textbf{p}$ calculation. Black dashed lines indicate the energetic position of the Fermi energy, $E_F$, for $V_g = -1.0$ and $-2.0$~V. Red and green dashed lines correspond to position of the Fermi energies of the red and green Hall resistance traces of Fig.\ A.
The Landau level crossing points are marked by arrows of the same color.
\bigskip

\textbf{Fig. 2} Bulk and edge state spectrum of the four band
Dirac model described by Eq.\ \ref{dirac} in the presence of an
external orbital magnetic field. (A) The bulk Landau levels in the
normal regime. (B) The bulk Landau levels in the inverted regime. A
pair of low lying Landau levels cross at a finite magnetic field
$B^c_\perp$. The crossing divides the phase diagram of gate voltage and
magnetic field into four regimes, labelled (i-iv) in the figure.
(C) The low lying bulk and edge state energies as a function of the
centers of the Landau orbitals in the normal regime. (D) The low
lying bulk and edge state energies as a function of the centers of
the Landau orbitals for the inverted regime, where the Fermi
energy lies in between the two bulk inverted Landau levels. The
Fermi energy crosses the Landau levels, giving rise to the one pair
of counter-propagating edge states. When the magnetic field is
increased, the two lowest lying bulk Landau levels approach each
other, and they cross the Fermi energy in different order,
depending on the value of the gate voltage. The crossing of the
Fermi energy by one of the Landau levels gives rise to either the
$n-$ or the $p$-type QHE for the cases shown in figures (E) and
(F).
\bigskip

\textbf{Fig. 3} Crossing field, $B^c_\perp$, (red triangles) and energy gap, $E_g$, (blue open dots) as a function of QW width $d$ resulting from an eight band $\textbf{k}\cdot\textbf{p}$ calculation. For well widths larger than 6.3 nm the QW is inverted and a mid-gap crossing of Landau levels deriving from the $HH1$ conductance and $E1$ valence band occurs at finite magnetic fields. The experimentally observed crossing points are indicated by open red squares.
The inset shows the energetic ordering of the QW subband structure as a function of QW width $d$. [taken from Ref. \cite{pfeuffer-jeschke2000}]
\bigskip

\textbf{Fig. 4}
The longitudinal four-terminal resistance, $R_{14,23}$, of various normal ($d = 5.5$~nm) (I) and inverted ($d= 7.3$ nm) (II, III, and IV) QW structures as a function of the gate voltage measured for $B=0$ T at $T=30$~mK. The device sizes are $(20.0 \times 13.3)$ $\mu$m$^2$ for device I and II, $(1.0 \times 1.0)$ $\mu$m$^2$ for device III, and $(1.0 \times 0.5)$ $\mu$m$^2$ for device IV. The inset shows $R_{14,23}(V_g)$ of two samples from the same wafer, having the same device size (III) at 30 mK (green) and 1.8 K (black) on a linear scale.
\bigskip

\textbf{Fig. 5}
Four-terminal magnetoconductance, $G_{14,23}$, in the QSH regime as a function of tilt angle between the plane of the 2DEG and applied magnetic field for a $d = 7.3$ nm QW structure with dimensions $(L\times W) = (20 \times 13.3)$ $\mu$m$^2$ measured in a vector field cryostat at 1.4 K.

\newpage

\includegraphics[width=14cm]{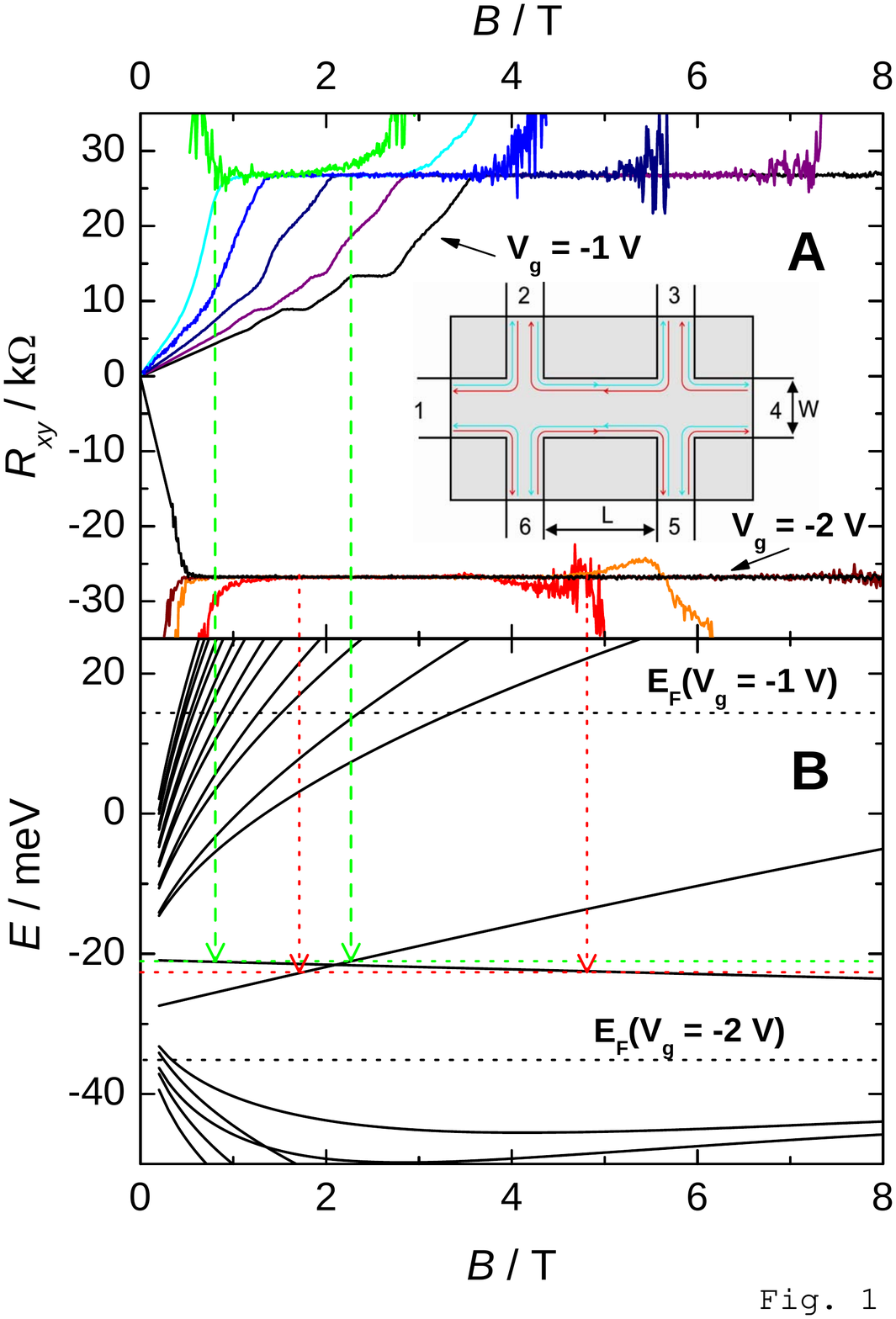}

\begin{center} {\Large Fig.\ 1}
\end{center}
\newpage

\includegraphics[width=14cm]{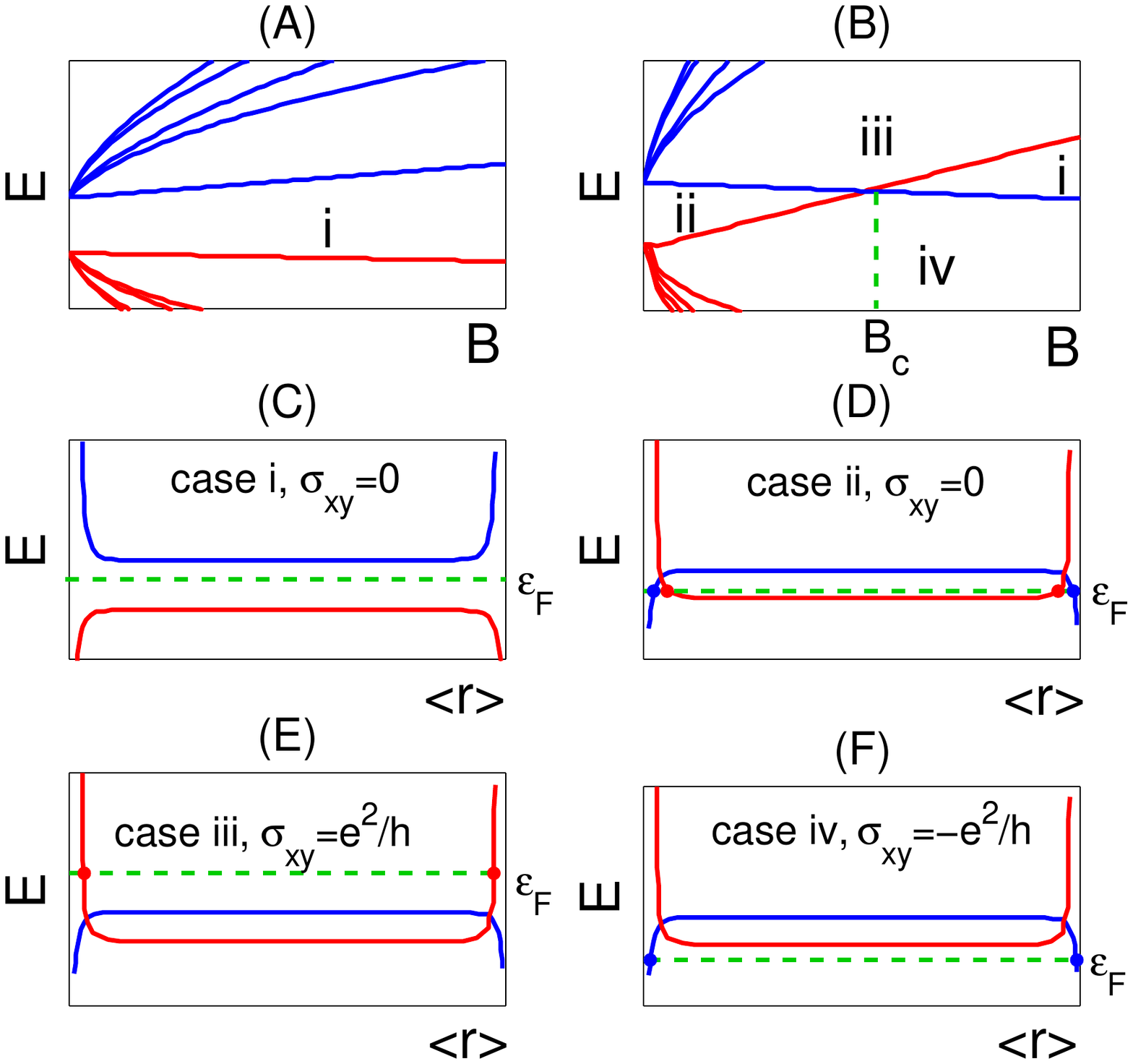}

\begin{center} {\Large Fig.\ 2}
\end{center}

\newpage

\includegraphics[width=16cm]{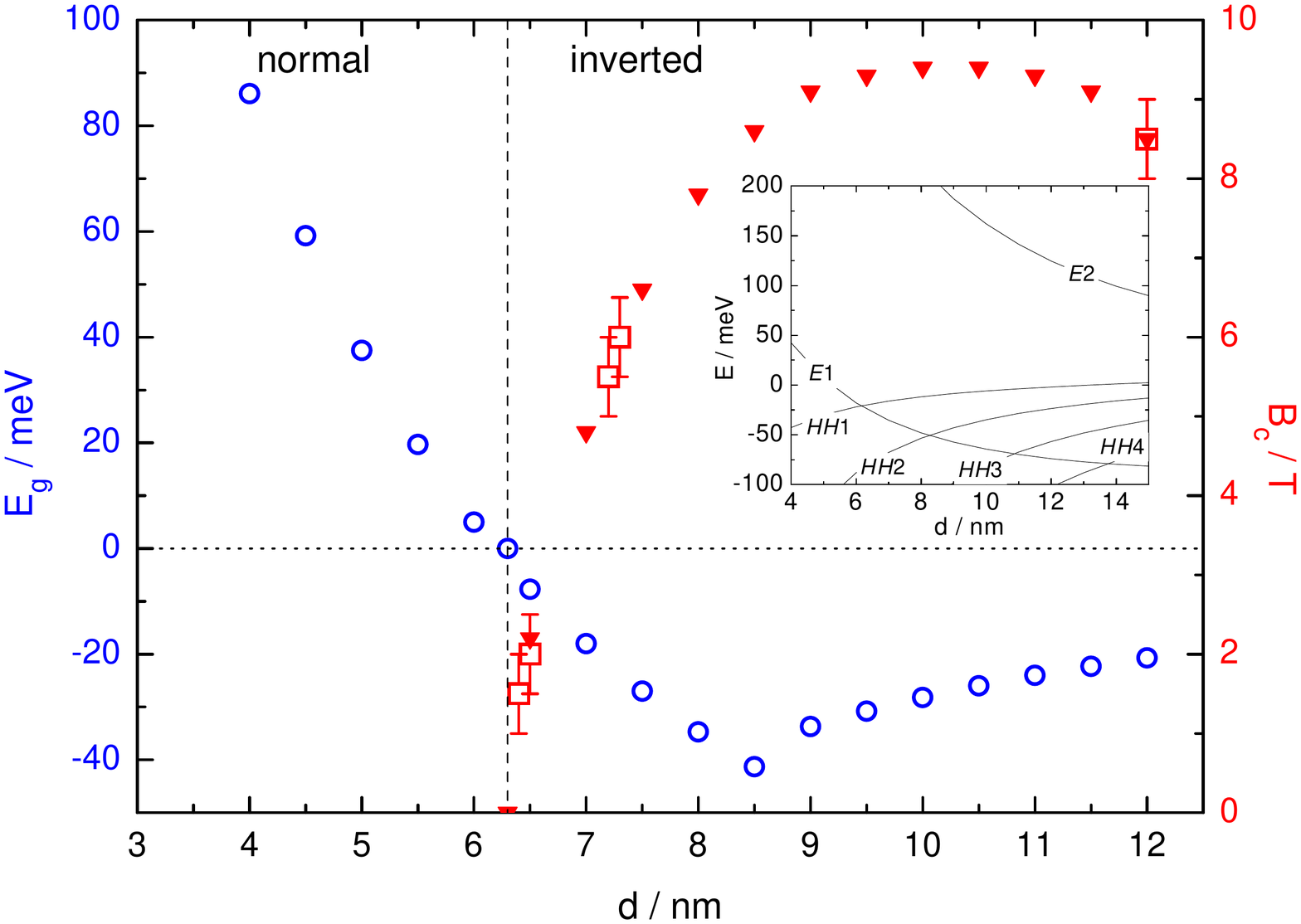}
\vfill
\begin{center} {\Large Fig.\ 3}
\end{center}

\newpage

\includegraphics[width=16cm]{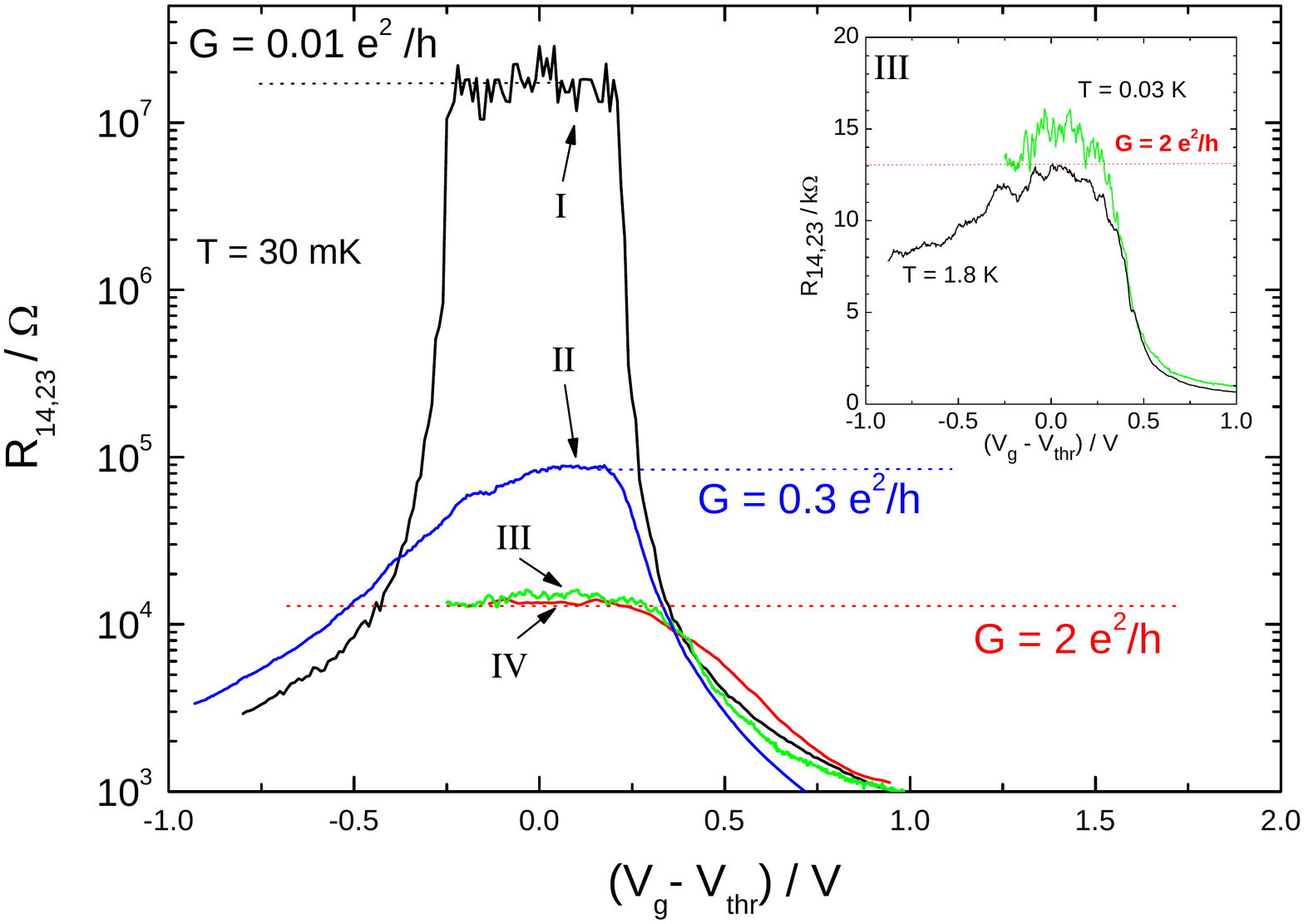}
\vfill
\begin{center} {\Large Fig.\ 4}
\end{center}
\newpage

\includegraphics[width=16cm]{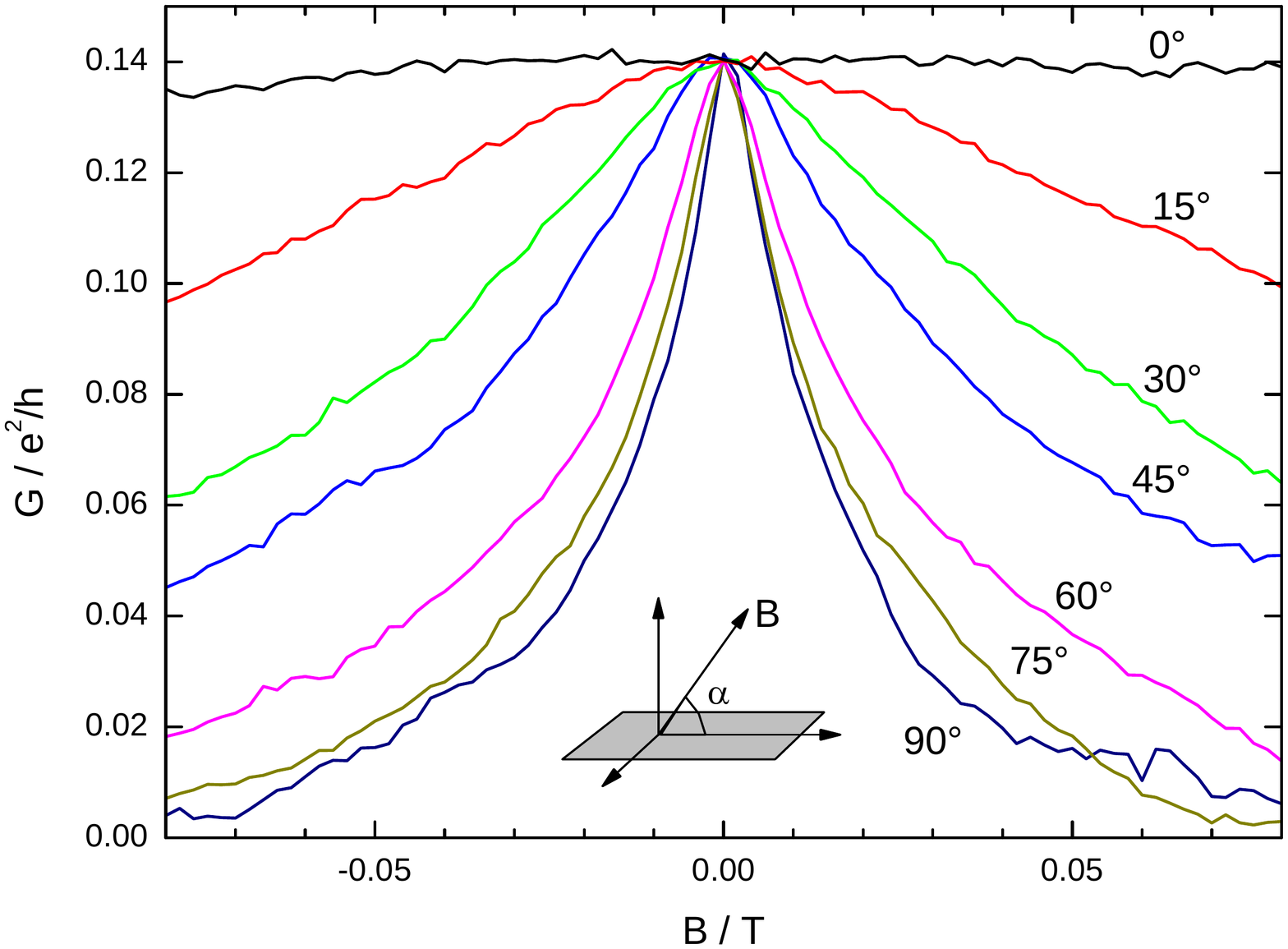}
\vfill
\begin{center} {\Large Fig.\ 5}
\end{center}

\end{document}